\documentclass[twocolumn,preprintnumbers,amsmath,amssymb]{revtex4}

\usepackage{graphicx}
\usepackage{dcolumn}
\usepackage{bm}

\begin{document}

\title{A Sensor Based on Extending the Concept of Fidelity to
Classical Waves}

\author{Biniyam Tesfaye Taddese\textsuperscript {1}}
\author{James Hart\textsuperscript {2}}
\author{Thomas M. Antonsen\textsuperscript {1, 2}}
\author{Edward Ott\textsuperscript {1, 2}}
\author{Steven M. Anlage\textsuperscript {1, 2}}
\affiliation{Department of Electrical and Computer Engineering, University of Maryland, College Park, Maryland 20742-3285, USA\textsuperscript {1}, Department of Physics, University of Maryland, College Park, Maryland 20742-4111, USA\textsuperscript {2}
}

\date{\today}

\begin{abstract}
	We propose and demonstrate a remote sensor scheme by applying the quantum mechanical concept of fidelity loss to
classical waves.  The sensor makes explicit use of time-reversal invariance and spatial reciprocity in a wave chaotic
system to sensitively and remotely measure the presence of small perturbations. The loss of fidelity is measured through
a classical wave-analog of the Loschmidt echo by employing a single-channel time-reversal mirror to rebroadcast a probe
signal into the perturbed system.  We also introduce the use of exponential amplification of the probe signal to
partially overcome the effects of propagation losses and to vary the sensitivity.
\end{abstract}
\maketitle

	Many sensor technologies are based on measurement of the disturbance of waves broadcast to and received from a remote region (e.g., ultrasonic sensors, radar, sonar, seismometers, etc.).  In most cases the sensors work best when there is a single path of propagation from the source to the target to the receiver.   In some cases there are multiple paths of propagation, and these can confound the sensor.  In the extreme case of an enclosure in which the trajectories of waves are chaotic (that is the trajectories depend sensitively on initial conditions and extend throughout the enclosure), the conventional approach of analyzing the returned signal assuming that it has propagated along known, predetermined trajectories fails.  This is the regime of wave/quantum chaos \cite{Stockmann}.

For insight into how chaos can enhance the operation of wave-based sensors, we turn to quantum mechanics for inspiration.
Quantum fidelity is a measure of how sensitive the dynamics of a time reversal invariant quantum mechanical system is to
small perturbations of its Hamiltonian.  It can be defined as follows.  A system is prepared in a given initial state
$|\Psi(0)\rangle$, propagated forward in time under an unperturbed time reversible Hamiltonian $H$ to some time $t$,
$|\Psi(t)\rangle=U(t)|\Psi(0)\rangle$  where $U(t)=exp(-iHt/\hbar)$ is the time evolution operator.  At that time the
evolution is stopped and the system is propagated backward in time under a perturbed Hamiltonian $H+\delta H$ to create
a state $U'(-t)U(t)|\Psi(0)\rangle$ where $U'(-t)=exp[i(H+\delta H)t/\hbar]$.  The overlap of this forward and backward
propagated state with the initial state is known as the fidelity, $f_{\delta H}(t)=<\Psi(0)|U'(-t)U(t)|\Psi(0)>$.
The fidelity is unity in the absence of perturbations for any $H$ and $t$.  However, in the presence of perturbations
the fidelity will decay with $t$ at a rate depending on $H$ and the
perturbation.  Fidelity is also known as the Loschmidt echo \cite{Gorin06a}, and thus makes connection to spin-echo
experiments widely used in nuclear magnetic resonance \cite{Slichter}.

Our sensor exploits an analogous effect. If a wave signal is launched from an antenna located in a reciprocal enclosure with ports,
and all the signal power is captured at the ports, and the port signals are time reversed and re-injected into the ports,
then a time reversed replica of the original signal will reassemble at the location of the antenna.  Remarkably, we shall
see that this reassembly process can be effective even if there is loss of signal, and even if the enclosure has chaotic
trajectories. The reassembly is degraded if the enclosure is perturbed between the original broadcast of the signal
and the time reversal and re-injection of the signal at the collecting ports.

An alternative, but equivalent, definition of fidelity, which we label the ``propagation comparison'' is simply to calculate the overlap between states at time $t$ that have been propagated forward from the same initial state by both the perturbed and unperturbed Hamiltonians. While the two definitions of fidelity are mathematically equivalent, their implementations can be quite different.

The ``propagation comparison'' concept of scattering fidelity has already been applied to classical wave
systems \cite{Gorin06b, Schafer05}. However, the repetitive collection of long complicated signals, and the cross-correlation
of them against a baseline signal, are both expensive in terms of storage and computational overhead. On the other
hand, the ``Loschmidt echo'' definition of fidelity now shows considerable promise with the development of `time-reversal
mirrors' for classical waves in acoustics \cite{Fink96, Fink00} and electromagnetics \cite{Lerosey, Anlage}.  Such mirrors
collect and record a propagating wave as a function of time, and at some later time propagate it in the opposite direction
in a time-reversed fashion.  In general it is not possible to mirror all waves in this manner.  However, this problem is
mitigated considerably in the case of a system with classically chaotic ray dynamics, where a
single-channel time-reversal mirror can very effectively approximate the conditions required to implement
the ``Loschmidt echo'' definition of fidelity \cite{Draeger97, Anlage}. In this paper we develop a sensor paradigm for
classical-wave-based sensors by measuring the scattering fidelity of a ray-chaotic system through the coherent
time-reversed reconstruction of an excitation pulse.

Fidelity has been shown to be a very sensitive measure of changes in the Hamiltonian in quantum systems \cite{Gorin06a}.  However, quantum systems have no dissipation, whereas classical wave scattering systems often have significant dissipation. One can think of the pulse propagation process in terms of waves traveling on a large number of semi-classical paths through the environment connecting the source and receiver, each path bouncing many times off of objects or boundaries.  These many ray paths act in parallel and each independently carries information \cite{Beenakker97}.  It is the coherent superposition of waves that travel along all of these different trajectories that leads to the sharp and dramatic re-construction of a time-reversed version of the original pulse.  The effect of uniform dissipation in the medium is to add uniform attenuation to the waves propagating on each of these trajectories or scattering channels, but not to change their phase.  The effect of a perturbation is to modify the phases (and amplitudes) of a finite subset of these ray paths, resulting in a reduced coherent re-construction.  Here we demonstrate a method to partially compensate for dissipation in classical wave scattering systems and demonstrate the efficacy of the Loschmidt echo for detecting small changes in scattering in the presence of dissipation.

	Both acoustic and electromagnetic classical waves have been employed for this work, but here we will focus on the acoustic case alone.  Acoustic time-reversal mirrors \cite{Fink96} have been used for, among other things, sound focusing \cite{Yon93}, and improved acoustic communications in air \cite{Candy06}.

\begin{figure}
\includegraphics[width=85mm]{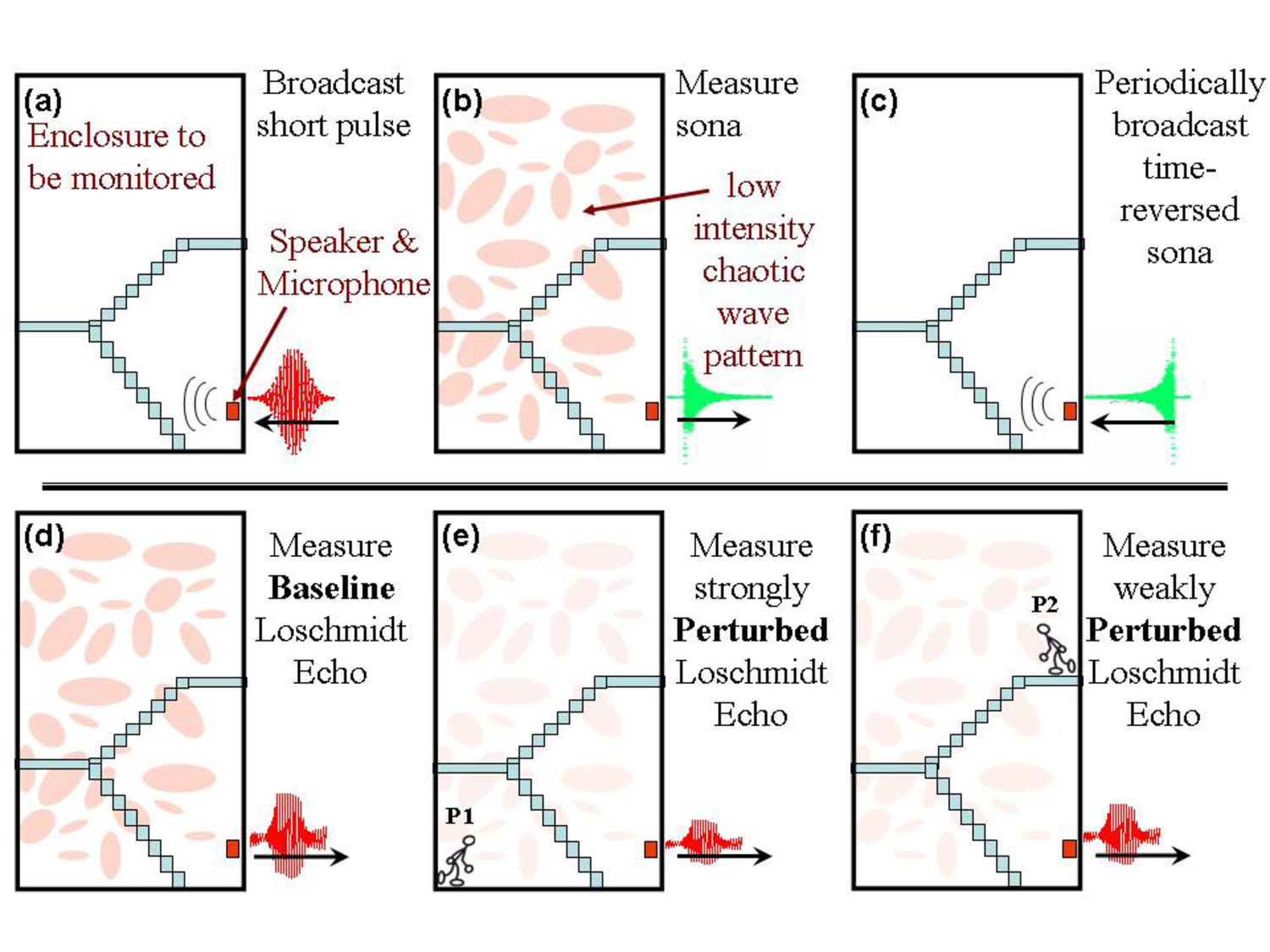}
\caption{\label{fig:Fig1} (Color online) Schematic diagram of stairwell and operation of the acoustic Loschmidt echo sensor.  }
\end{figure}

The ray-chaotic enclosure in which the sensor is demonstrated is a 2-story-tall enclosed stairwell, roughly
6 m deep x 2.5 m wide x 6.5 m tall, containing stairs with an intermediate landing (see Fig.~\ref{fig:Fig1}).
Acoustic waves are launched into this quiescent air-filled enclosure using a standard audio speaker, and measured with a
Samson C01U microphone.  The speaker and microphone operate over the range of about 30 Hz to 15 kHz and are connected
to a computer located outside the enclosure.  Various objects are introduced into and/or removed from the enclosure to
test the sensitivity of the wave dynamics to perturbations.

	The acoustic time-reversal mirror operating in a Loschmidt echo configuration works as follows.  A short
Gaussian-in-time pulse of a fixed carrier frequency tone is generated by the computer and broadcast into the acoustic
enclosure through the audio speaker (Fig. 1(a)). Typical carrier frequency and duration of the pulse are $f =$ 7 kHz
and 2 ms, respectively, and the waves have a wavelength of $\lambda \sim$ 5 cm, which is much smaller than the enclosure
size.  The character of the pulse generated by the speaker is independently measured in an anechoic enclosure (Fig. 2(a)).
The time-dependent `sona' signal (Figs. 1(b), 2(b)) is measured by the microphone at a separate location many wavelengths
away from the source.  This signal is amplified, digitized, and recorded by the computer.  The time-reversed sona signal is formed in the computer and launched from the speaker into the unperturbed room (Fig. 1(c)). This is done
without exchanging the positions of the speaker and microphone, hence it is assumed that spatial reciprocity holds. The
waves eventually arrive at the microphone and re-construct in a time-reversed approximation to the original pulse, even when the enclosure is perturbed (Figs. 1(d-f), 2(c)).
Pulses are re-constructed before and after the cavity perturbation, and they are compared to each other to make an estimate of the scattering fidelity of the system.
Note that the sensor detects changes in the overall scattering environment and does not directly reveal the location or volume of the
scattering perturbation.
\begin{figure}
\includegraphics[width=85mm]{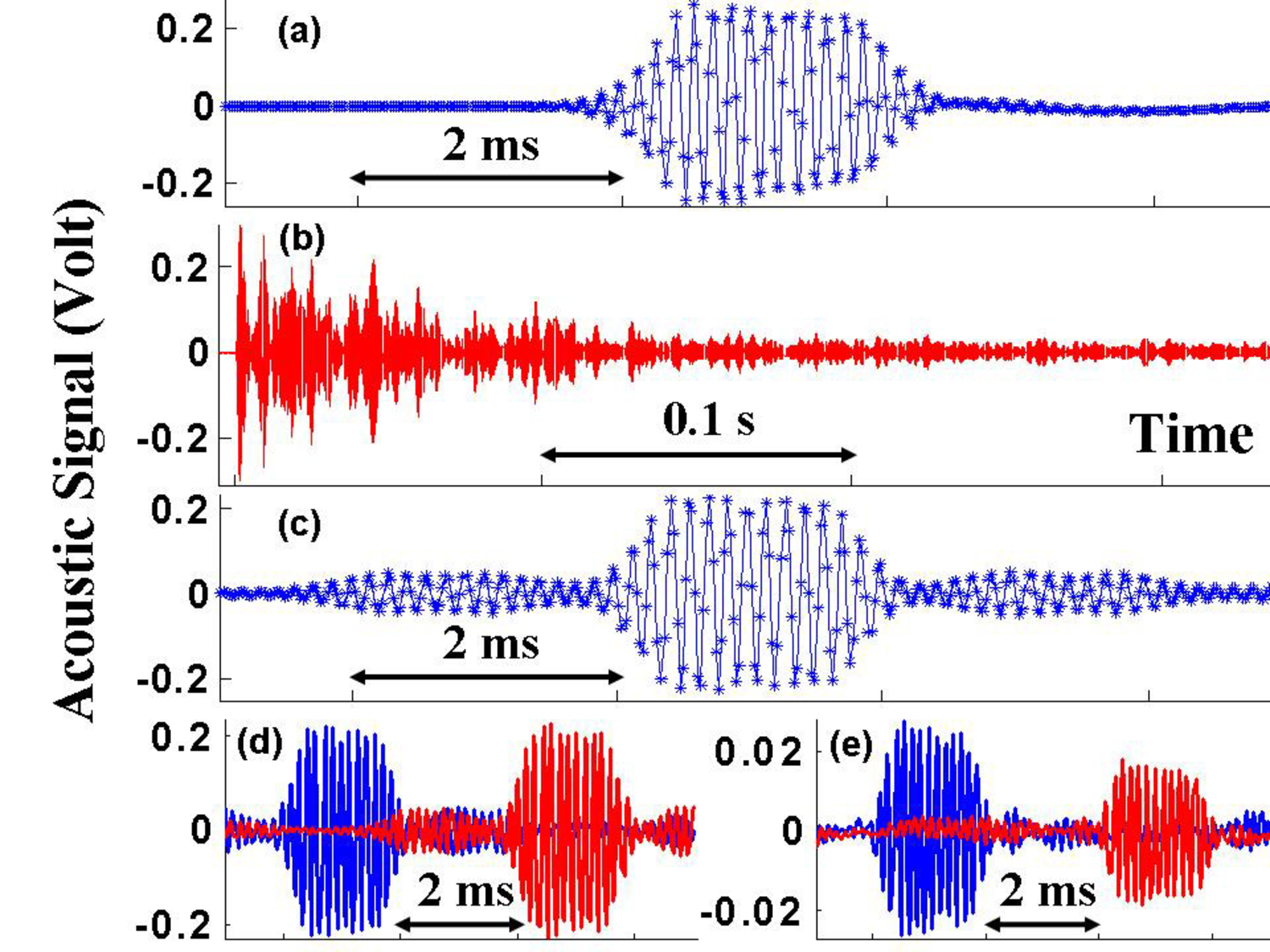}
\caption{\label{fig:Fig2} (Color online)  (a) The original acoustic pulse recorded with a 36 dB pre-amplification in an
anechoic chamber.  (b)  The sona signal recorded in the stairwell with 48 dB pre-amplification. (c) The LE
(time reversed pulse) recorded in the stairwell with 36 dB
pre-amplification.  (d) Comparing LE pulses with no amplification of the sona before (blue/left) and after (red/right)
distant perturbation at P2.
(e) Comparing LE pulses with exponential amplification of the sona before (blue/left) and after (red/right) distant
perturbation at P2.}
\end{figure}

The reconstructed pulse is not an exact time-reversed duplicate of the original pulse.  There are a number of reasons
for this including: i) use of a finite time-window when recording the sona signal, and ii) dissipation in the system.
Concerning the effect of i), it was shown \cite{Draeger99} that the quality of the time-reversal focusing is dependent
on the size of the time-reversal window.  In our experiment a $\sim 3 s$-long window is not sufficient to capture the
entire sona wave dynamics required to reconstruct the pulse.

With respect to the effect of ii), it should be
noted that the acoustic enclosure has losses associated with
propagation through the air and absorption in the walls,
floor, ceiling, and stairs.  The loss parameter of the cavity
$(\alpha \sim 1200)$, defined as the ratio of the typical 3-dB bandwidth of
the resonance modes to the mean spacing between
eigenfrequencies \cite{Hemmady}, implies that the modes are
strongly overlapping.  This also results in uniform loss of
information and a degradation of the echo.
In spite of the significant loss and short sona recording
window, we still observe (Fig. 2) good pulse reconstruction.

The effect of uniform dissipation in the medium is to add an exponential decay to
the measured sona signals.  This limits the sensitivity of
the Loschmidt echo (LE) to perturbations of the scattering enclosure.
We explored the effects of dissipation, perturbation strength, and measurement limitations by producing a variety of perturbations to
the acoustic enclosure and measured their effects on the
LE.  First a baseline Loschmidt echo (BLE)
is measured immediately after the unperturbed sona signal is
collected (Fig. 1(d)). Next a perturbation is made to the scattering
environment, and a perturbed Loschmidt echo (PLE) is measured (Figs. 1(e,f)).
Comparison between the unperturbed and perturbed echoes can be
done either by cross-correlation, or by simply comparing the
peak-to-peak amplitudes (PPA) of the reconstructed pulse signals.
When a perturbing object (50 cm x 30 cm x 15 cm cloth backpack, inducing a fractional enclosure volume change of 2 10$^{-4}$) is
added to the acoustic enclosure on the ground floor about 2
meters from the speaker and microphone (P1 in Fig. 1(e)), there
is an 8\% drop in the PPA of the PLE
compared to the BLE.  The statistical fluctuation of the PPA observed in
control experiments is about 2\%.  However, if the same perturbing object is placed on the second floor of the
enclosure (P2 in Fig. 1(f)), about 5 meters away with no line-of-sight propagation path from the microphone or speaker,
the PPA of the BLE and PLE are the same within statistical fluctuations (Fig 2(d)).

The LE is insensitive to perturbations of the scattering environment at locations where the scattered waves suffer
significant attenuation before reaching the detector. To partially overcome the loss limitations of the LE, we have
applied an exponential amplification to the measured sona signal before time-reversal.  Ideally the amplification will
substantially remove the decay brought on by the dissipative wave propagation, thus mitigating effect ii) mentioned above.
  However, the finite recording dynamic range of the microphone limits the duration of the exponential amplification.  In addition, the amplification
must be turned off smoothly to prevent additional frequency components from entering the time-reversed sona signal and
corrupting the reconstructed pulse.  The following generic amplification function $A(t)$ has been employed:
\begin{equation}
A(t)=\left[
 1-4(\frac{t}{W})^{6}+3(\frac{t}{W})^{8} \right]
 exp(\frac{Ft}{\tau}), \hspace{5 mm} (0 \leq t \leq W)
\end{equation}
where $t$ is time, $W$ is the width in time of the amplifying
window, $F$ is the exponent parameter, and $\tau$ is the measured
$1/e$ decay time of the enclosure.  The polynomial smoothly
turns off the amplifying function at $t = W$.  One expects that an exponent parameter $F = 2$ will compensate for the effect of attenuation upon forward and backward propagation.

    The experiments discussed above were repeated with the
exponential amplification applied to the measured sona signal.
The values of $W$ and $F$ were systematically varied to maximize the sensitivity of the LE to particular perturbations.
In this case the nearby perturbation (backpack placed at P1 in Fig. 1(e)) resulted
in a 40\% change in PPA of the PLE compared
to the BLE, using values of $W = 0.8 s$ and $F = 2$.  A distant non-line-of-sight perturbation (backpack placed at P2 in Fig. 1(f)) was
now clearly
detected, resulting in a 30\% change between the PLE and BLE (see Fig. 2(e)) using values of $W = 0.9 s$ and $F = 3$.
In general, non-line-of-sight perturbations are only resolved using the exponential amplification algorithm.
The sensor operates in real-time, producing LE pulses at a rate limited only by the decay time of acoustic energy
in the enclosure. As such, it is best suited for detecting a change in the environment after it has been returned to its nominal initial state. The electromagnetic implementation can operate $\sim$$10^{6}$ times faster and is better suited for dynamic sensing.

To utilize the exponential amplification algorithm to improve the LE measurement one must first calibrate the system
by systematically varying the $W$ and $F$ parameters for a given decay time $\tau$ of the enclosure and characteristic
perturbation of the scattering environment.  By varying the parameters it is possible to customize the sensor to detect
certain types of perturbations at certain locations.  These variations of the amplification parameters can be executed
dynamically so that the sensor systematically explores the enclosure tuned to different types of
perturbations.  Comparing the LE and propagation comparison methods, we note that measurement of the LE can be done
with a simple circuit enabling immediate detection of a change, whereas in propagation comparison a computationally intensive cross-correlation
must be computed first.  Finally we have found that both detection methods benefit from exponential amplification, Eq. (1).

In conclusion, we have developed a sensor paradigm that makes use of chaotic ray dynamics, as well as time-reversal invariance and spatial reciprocity properties of wave propagation, to sensitively measure small perturbations to wave scattering systems.  The sensor makes use of a Loschmidt echo (scattering fidelity decay) experiment applied to classical waves to measure the sensitivity of a system's dynamics to small perturbations.

This work is supported by ONR MURI grant N000140710734, AFOSR grant FA95500710049, and the Maryland Center for Nanophysics and Advanced Materials.

\end{document}